\documentclass[11pt,oneside,letterpaper]{article}
\usepackage{geometry}                		%
\usepackage{graphicx}				%
\usepackage{amssymb}
\usepackage{physics}
\usepackage[colorlinks=false, urlcolor=blue, linkcolor=red]{hyperref}
\usepackage{cite}
\usepackage[symbol]{footmisc}

\title{\bf Velocity, temporal and generalized Kirchhoff gauges}
\author
{Kuo-Ho Yang \\
Department of Engineering and Physics, St.~Ambrose University,\\ 
Davenport, IA 52803 \\
E-mail: yangkuoho@sau.edu\vspace{3mm}\\
Robert D. Nevels\\
Department of Electrical and Computer Engineering, Texas A\ \&\ M University,\\
College Station, TX 77843
}
\date{}			

\begin{document}
\maketitle

\begin{abstract}
We derive the dyadic Green's function for the generalized velocity gauge in the time domain with an arbitrary parameter $v$ and show that it reduces to the Lorenz and the Coulomb gauge Green's functions in the limit as $v$ goes to the speed of light $c$ and to infinity, respectively. We also derive the Green's function for the generalized Kirchhoff gauge in the frequency domain and show that it reduces to the results of the Coulomb and the temporal gauges as special cases.   Together the generalized velocity and Kirchhoff gauges encompass most ${\bf A}/{\Phi}$ gauges in electromagnetics.
\end{abstract}

\section{Introduction}

In a previous paper~\cite{Yang-gauge-2005}, Yang reviewed a new approach to classical electromagnetic potentials and gauge transformations. He discussed the Coulomb and the Lorenz gauges, some serious defects of the Helmholtz theorem, and a generalized class of gauges called the velocity gauge with a parameter $v$, of which the Coulomb gauge ($v = \infty$) and the Lorenz gauge ($v = c$) are special cases. Most importantly, he obtained analytic solutions for the potentials, enabling us to visualize how the potentials propagate from physical charge and current densities. In the velocity gauge with $v \ne c$, the scalar potential propagates at speed $v$ and the vector potential has one component that propagates at this speed and another that always propagates at the finite speed $c$.   In an earlier paper, Yang and Kobe\cite{Yang-Kobe-gauge-1986} first investigated the effects of advanced potentials ($v < 0$) and faster-than-$c$ retarded potentials ($v > c$).  Using the propagation properties of these potentials, they established a link between the violation of gauge invariance and the violation of causality and relativity.

Throughout the long history of gauges, relevant work has been published widely in the literature \cite{Jackson-Okun-2001}. One series of extensive studies focused primarily on the Coulomb gauge \cite{Mich-Nevels-1988, Mich-Nevels-Zheng-1990, Zheng-Mich-Crow-Nevels-1990, Nevels-Crowell-1990, Rynne-Smith-Nevels-1991, Nevels-Wu-1992}. Some contributions in this research include analytic solutions for the Coulomb gauge dyadic Green's function in the frequency and time domains, the analytic and numerical solutions for the potentials in the Coulomb and the Lorenz gauges for various boundary conditions, and the computation of the fields and current densities from these solutions.

In this paper, we show that the velocity gauge integral equation and associated Green's function derived following Ref.~\cite{Yang-gauge-2005} are consistent with the results in Refs.\cite{Mich-Nevels-1988, Mich-Nevels-Zheng-1990, Zheng-Mich-Crow-Nevels-1990, Nevels-Crowell-1990, Rynne-Smith-Nevels-1991, Nevels-Wu-1992}, thereby expanding our knowledge of electromagnetic potentials. In particular, we will derive the Green's function for the velocity gauge in the time domain and show that it reduces to the frequency domain Lorenz and Coulomb\cite{Nevels-Crowell-1990} gauge expressions when $v$ goes to the speed of light $c$ and to infinity, respectively. We also derive the Green's function for the generalized Kirchhoff gauge in the frequency domain and show that it reduces to the Green's function for the temporal and the Coulomb gauges when the parameter $v$ goes to zero and infinity, respectively.

\section{Green's functions for velocity, temporal and generalized Kirchhoff gauges}

\begin{flushleft}
{\bf A. The velocity gauge}
\end{flushleft}

We start with the velocity gauge in terms of the parameter $v$ given in Eqn. (3.4) of Ref.~\cite{Yang-gauge-2005},
\begin {equation}
\grad \cdot {\bf A}^{(v)}({\bf r}, t) = - {c \over v^2} {\partial \Phi^{(v)}({\bf r}, t) \over \partial t},
\label{eq-A1}
\end{equation}
which by virtue of Maxwell’s equations leads to the scalar and vector potential equations:
\begin{equation}
\left( \grad^2 - {1 \over v^2}{\partial^2 \over \partial t^2} \right) \Phi^{(v)}({\bf r}, t) = - 4\pi \rho({\bf r}, t),
\label{eq-A2}
\end{equation}
\begin{equation}
\left( \grad^2 - {1 \over c^2}{\partial^2 \over \partial t^2} \right) {\bf A}^{(v)} ({\bf r}, t)
= - {4\pi \over c}{\bf J}({\bf r}, t) - c\left( {1 \over v^2} - {1 \over c^2} \right)  \grad \left({\partial \Phi^{(v)}({\bf r}, t) \over \partial t}\right).
\label{eq-A3}
\end{equation}
Here we use the notation of Ref.~\cite{Yang-gauge-2005} whereby $(v)$ used as a superscript represents the velocity gauge and as a parameter the values of $v$ distinguish the various potentials. When $v = c$, it reduces (\ref{eq-A1}) to the Lorenz gauge, and when $v = \infty$ (\ref{eq-A1}) becomes the Coulomb gauge. The solution to (\ref{eq-A2}) is
\begin{equation}
\Phi^{(v)}({\bf r}, t) = \int G({\bf r}, t |v| {\bf r}', t') \rho({\bf r}', t') d^3r'dt',
\label{eq-A4}
\end{equation}
where the $v$-propagating Green's function,
\begin{equation}
G({\bf r}, t |v| {\bf r}', t') = {\delta(t - {|{\bf r} - {\bf r}'| \over v} - t') \over |{\bf r} - {\bf r}'|},
\label{eq-A5}
\end{equation}
is a solution to the equation,
\begin{equation}
\left( \grad^2 - {1 \over v^2}{\partial^2 \over \partial t^2} \right) G({\bf r}, t |v| {\bf r}', t') 
= - 4 \pi \delta({\bf r} - {\bf r}') \delta(t - t'),
\label{eq-A6}
\end{equation}
and $\delta$ is the Dirac delta-function.

In Ref.~\cite{Yang-gauge-2005}, Eqn. (3.17) it was shown that the velocity-gauge vector potential can be placed in the form
\begin{equation}
{\bf A}^{(v)} ({\bf r}, t) = {\bf A}^{(L)} ({\bf r}, t) + c \grad \int^t \left[ \Phi^{(L)}({\bf r}, t'') - \Phi^{(v)}({\bf r}, t'') \right] dt''.
\label{eq-A7}
\end{equation}
Here the gauge superscript $( L )$ denotes that ${\bf A}^{(L)}$ and $\Phi^{(L)}$ are the Lorenz gauge potentials. Eqn.~(\ref{eq-A4}) is also the solution to the Lorenz gauge with $\Phi^{(v)}({\bf r}, t) \to \Phi^{(L)}({\bf r}, t)$ and 
$G({\bf r}, t |v| {\bf r}', t') \to G({\bf r}, t |c| {\bf r}', t')$.

To derive an analytic expression for the dyadic Green's function $\bar{{\bf G}}^{(v)}({\bf r}, t || {\bf r}', t')$ for the vector potential,  
\begin{equation}
{\bf A}^{(v)}({\bf r}, t) = {1 \over c} \int \bar{{\bf G}}^{(v)}({\bf r}, t || {\bf r}', t') \cdot {\bf J}({\bf r}', t') d^3r'dt',
\label{eq-A8}
\end{equation}
we must first convert the two scalar potentials in Eqn.~(\ref{eq-A7}) with the charge density as the source into an expression with the current density as the source. The conversion can be done through (\ref{eq-A4}) and the continuity equation as follows:
\begin{eqnarray}
c \grad \int^t \left[ \Phi^{(L)}({\bf r}, t'') - \Phi^{(v)}({\bf r}, t'') \right] dt''
\qquad \qquad \qquad \qquad \qquad \qquad \qquad \qquad \qquad \quad 
\nonumber
\\
= c \grad \int^t dt'' \int^{t''} d\tau \left[ {\partial \Phi^{(L)}({\bf r}, \tau) \over \partial \tau} - {\partial \Phi^{(v)}({\bf r}, \tau) \over \partial \tau}  \right]
\qquad \qquad \qquad \qquad \qquad \qquad \qquad \ 
\nonumber
\\
= c \grad \int^t dt'' \int^{t''} d\tau \left[ G({\bf r}, \tau |c| {\bf r}', t') - G({\bf r}, \tau |v| {\bf r}', t')   \right] {\partial \rho({\bf r}', t') \over \partial t'} d^3r'dt'
\qquad \qquad \qquad
\nonumber
\\
= - c \grad \int^t dt'' \int^{t''} d\tau \left[ G({\bf r}, \tau |c| {\bf r}', t') - G({\bf r}, \tau |v| {\bf r}', t')   \right] \left[ \grad' \cdot {\bf J}({\bf r}', t') \right]d^3r'dt'
\qquad \quad
\nonumber
\\
= - c \int \left\{ \grad \grad \int^t dt'' \int^{t''} d\tau \int \left[ G({\bf r}, \tau |c| {\bf r}', t') - G({\bf r}, \tau |v| {\bf r}', t')   \right] \right\} \cdot {\bf J}({\bf r}', t')d^3r'dt'.
\label{eq-A9}
\end{eqnarray}
Thus, we obtain the complete Green's function in the velocity gauge,
\begin{eqnarray}
\bar{{\bf G}}^{(v)}({\bf r}, t || {\bf r}', t') = G({\bf r}, t |c| {\bf r}', t')\bar{{\bf I}}
\qquad \qquad \qquad \qquad \qquad \qquad \qquad \qquad \quad
\nonumber
\\
\qquad \qquad \qquad
- c^2 \grad \grad \int^t dt'' \int^{t''} \left[ G({\bf r}, \tau|c|{\bf r}', t') - G({\bf r}, \tau|v|{\bf r}', t') \right] d\tau,
\label{eq-A10}
\end{eqnarray}
where $\bar{{\bf I}}$ is the identity matrix.

We now show that Eqn.~({\ref{eq-A10}) is equivalent to another expression in Ref.~\cite{Yang-gauge-2005}, Eqn. (3.32):
\begin{equation}
\bar{{\bf G}}^{(v)}({\bf r}, t || {\bf r}', t') = G({\bf r}, t |c| {\bf r}', t')\bar{{\bf I}}
 + {v^2 - c^2 \over 4\pi v^2} \grad \grad G({\bf r}, t|c|v|{\bf r}', t'),
\label{eq-A11}
\end{equation}
where
\begin{equation}
G({\bf r}, t|c|v|{\bf r}', t') = \int G({\bf r}, t |c| {\bf r}'', t'')G({\bf r}'', t'' |v| {\bf r}', t')d^3r''dt'',
\label{eq-A12}
\end{equation}
is a solution to the equation,
\begin{equation}
\left( \grad^2 - {1 \over v^2}{\partial^2 \over \partial t^2}\right) \left( \grad^2 - {1 \over c^2}{\partial^2 \over \partial t^2}\right) G({\bf r}, t|c|v|{\bf r}', t')
= (-4\pi)^2 \delta({\bf r} - {\bf r}') \delta(t - t'),
\label{eq-A13}
\end{equation}
in infinite free space with no boundary surfaces.

We evaluate the last term of Eqn.~(\ref{eq-A11}) using a method that is simpler than that of
direct integrations by Brown and Crothers\cite{Brown-Crothers-1989}, because of the presence of $\delta({\bf r} - {\bf r}')$ and $\delta(t - t')$ in Eqn.~(\ref{eq-A13}).  We first note that 
$\grad G({\bf r}, t|c|v|{\bf r}', t') = - \grad'G({\bf r}, t|c|v|{\bf r}', t')$ and 
$(\partial/\partial t)G({\bf r}, t|c|v|{\bf r}', t') = - (\partial/\partial t')G({\bf r}, t|c|v|{\bf r}', t')$.   We then differentiate the last term of Eqn.~(\ref{eq-A11}) twice with respect to time $t$ and use (\ref{eq-A6}) for the Green's function $G({\bf r}, t|v|{\bf r}', t')$ to
simplify the integration over $d^3r''$ and $dt''$ ,
\begin{eqnarray}
{\partial^2 \over \partial t^2} \left[{v^2 - c^2 \over 4\pi v^2} G({\bf r}, t|c|v|{\bf r}', t') \right]
= {c^2 \over 4\pi} \left[ \left( \grad'^2 - {1 \over v^2}{\partial^2 \over \partial t'^2}\right) - \left( \grad^2 
   - {1 \over c^2}{\partial^2 \over \partial t^2}\right) \right] G({\bf r}, t|c|v|{\bf r}', t')
\nonumber
\\
= {c^2 \over 4\pi} \int G({\bf r}, t|c|{\bf r}'', t'') \left[ \left( \grad'^2 - {1 \over v^2}{\partial^2 \over \partial t'^2}\right) G({\bf r}'', t''|v|{\bf r}', t') \right] d^3r''dt''
\qquad \qquad \qquad \qquad \qquad 
\nonumber
\\
 - {c^2 \over 4\pi} \int \left[ \left( \grad^2 - {1 \over c^2}{\partial^2 \over \partial t^2}\right) G({\bf r}, t|c|{\bf r}'', t'') \right] G({\bf r}'', t''|v|{\bf r}', t') d^3r''dt''
 \qquad \qquad \qquad \qquad 
 \nonumber
\\
=- c^2 [ G({\bf r}, t|c|{\bf r}', t') - G({\bf r}, t|v|{\bf r}', t') ].
 \qquad \qquad \qquad \qquad \qquad \qquad \qquad \qquad \qquad \qquad \quad \quad \ 
 \label{eq-A14}
\end{eqnarray}
Thus, our proof of the equivalence of Eqns.~(\ref{eq-A10}) and (\ref{eq-A11}) is completed.

We use ${\bf J}({\bf r},t) = {\bf J}_\omega ({\bf r}) e^{-i \omega t}$ and Eqn.~(\ref{eq-A5}) in Eqn.~(\ref{eq-A10}) to obtain the result~\cite{note-Fourier},
\begin{eqnarray}
\int \bar{{\bf G}}^{(v)}({\bf r}, t || {\bf r}', t') \cdot {\bf J}_\omega({\bf r}') e^{-i \omega t} d^3r'dt'
\qquad \qquad \qquad \qquad \qquad \qquad \qquad 
 \nonumber
\\
= e^{-i \omega t} \int \left\{ {e^{i(\omega/c)R} \over R}\bar{{\bf I}} 
  + {c^2 \over \omega^2} \grad \grad \left[ {e^{i(\omega/c)R} \over R} - {e^{i(\omega/v)R} \over R} \right]\right\} 
  \cdot {\bf J}_\omega ({\bf r}') d^3r',
 \label{eq-A15}
\end{eqnarray}
where $R = |{\bf r} - {\bf r}'|$.  Hence the $v$-gauge dyadic Green's function in the frequency domain is
\begin{equation}
 \bar{{\bf G}}_\omega^{(v)}({\bf r} || {\bf r}') = {e^{i(\omega/c)R} \over R}\bar{{\bf I}} 
   + {c^2 \over \omega^2} \grad \grad \left[ {e^{i(\omega/c)R} \over R} - {e^{i(\omega/v)R} \over R} \right].
\label{eq-A16}
\end{equation}
In the limit as $v$ goes to the speed of light $c$, (16) reduces to the Lorenz gauge Green's function:
\begin{equation}
\bar{{\bf G}}_\omega^{(L)}({\bf r} || {\bf r}') = {e^{i(\omega/c)R} \over R}\bar{{\bf I}}.
\label{eq-A17}
\end{equation}
Also, in the limit as $v$ goes to infinity, (16) reduces to the Coulomb gauge Green's function:
\begin{equation}
 \bar{{\bf G}}_\omega^{(C)}({\bf r} || {\bf r}') = {e^{i(\omega/c)R} \over R}\bar{{\bf I}} 
   + {c^2 \over \omega^2} \grad \grad \left[ {e^{i(\omega/c)R} \over R} - {1 \over R} \right],
\label{eq-A18}
\end{equation}
which agrees with the result in Ref.\cite{Nevels-Crowell-1990}.

\begin{flushleft}
{\bf B. Temporal gauge}
\end{flushleft}

The gauge condition and the potentials in the temporal gauge are ({\it e.g.}~Refs.~\cite{Yang-gauge-2005, Jackson-Okun-2001}):
\begin{equation}
{1 \over c} {\partial \over \partial t} \left[\grad \cdot {\bf A}^{(T)} ({\bf r}, t) \right] = - 4\pi \rho({\bf r}, t),
\label{eq-A19}
\end{equation}
\begin{equation}
\Phi^{(T)}({\bf r}, t) = 0,
\label{eq-A20}
\end{equation}
\begin{equation}
{\bf A}^{(T)}({\bf r}, t) = {\bf A}^{(L)}({\bf r}, t) + c \int^t \grad \Phi^{(L)} ({\bf r}, t'') dt'' = - c \int^t {\bf E}({\bf r}, t'') dt''.
\label{eq-A21}
\end{equation}
The Green's functions in the time and frequency domains can be obtained from Eqns.~(\ref{eq-A10}) and (\ref{eq-A16}) by dropping the $v$-propagating scalar potentials:
\begin{equation}
\bar{{\bf G}}^{(T)}({\bf r}, t || {\bf r}', t') = G({\bf r}, t |c| {\bf r}', t')\bar{{\bf I}}
  - c^2 \grad \grad \int^t d\tau \int^{\tau} G({\bf r},t'' |c|{\bf r}', t') dt'',
\label{eq-A22}
\end{equation}
\begin{equation}
 \bar{{\bf G}}_\omega^{(T)}({\bf r} || {\bf r}') = {e^{i(\omega/c)R} \over R}\bar{{\bf I}} 
   + {c^2 \over \omega^2} \grad \grad \left[ {e^{i(\omega/c)R} \over R} \right].
\label{eq-A23}
\end{equation}

\begin{flushleft}
{\bf C. Generalized Kirchhoff gauge}
\end{flushleft}

The generalized Kirchhoff gauge is defined by the gauge condition,
\begin{equation}
\grad \cdot {\bf A}^{(v{\rm K})} ({\bf r}, t) = {c \over v^2} {\partial \over \partial t} \Phi^{(v {\rm K})} ({\bf r}, t).
\label{eq-A24}
\end{equation}
When $v = c$, the generalized Kirchhoff gauge reduces to the original Kirchhoff gauge\cite{Jackson-Okun-2001} investigated recently by Heras\cite{Heras-2006}.  This gauge condition can be obtained from the velocity gauge condition by replacing $v^2$ in Eqn. (\ref{eq-A1}) by $(-v^2) = (\pm iv)^2$. Hence, the equations for the potentials are:
\begin{equation}
\left( \grad^2 + {1 \over v^2} {\partial^2 \over \partial t^2} \right) \Phi^{(v {\rm K})} ({\bf r}, t) = - 4 \pi \rho({\bf r}, t),
\label{eq-A25}
\end{equation}
\begin{equation}
\left( \grad^2 - {1 \over c^2} {\partial^2 \over \partial t^2} \right) {\bf A}^{(v {\rm K})} ({\bf r}, t) 
= - {4\pi \over c} {\bf J}({\bf r}, t) + c \left( {1 \over v^2} + {1 \over c^2} \right) \grad \left[ {{\partial \over \partial t}} \Phi^{(v {\rm K})} ({\bf r}, t) \right].
\label{eq-A26}
\end{equation}
If we use $\Phi^{(v {\rm K})} ({\bf r}, t)$ to denote the scalar potential (see the Appendix), the solution for the vector potential is:
\begin{equation}
{\bf A}^{(v {\rm K})} ({\bf r}, t) = {\bf A}^{(L)} ({\bf r}, t) + c \grad \int^t \left[ \Phi^{(L)} ({\bf r}, t'') - \Phi^{(v {\rm K})} ({\bf r}, t'') \right] dt'',
\label{eq-A27}
\end{equation}
where ${\bf A}^{(L)}$ and $\Phi^{(L)}$ are the potentials in the Lorenz gauge. Because of the boundary condition that the scalar potential $\Phi^{(v {\rm K})}$ goes to zero as $r$ goes to infinity, it is not as straightforward to write down the analytic solution of the Green's function for the generalized Kirchhoff gauge (see the Appendix). 

It is possible to derive the Green's function for the generalized Kirchhoff gauge in the frequency domain. We use $\rho({\bf r}, t) = \rho_{\omega} ({\bf r}) e^{-i\omega t}$, ${\bf J}({\bf r}, t) = {\bf J}_{\omega} ({\bf r}) e^{-i\omega t}$ and $\Phi^{(v {\rm K})}({\bf r}, t) = \Phi^{(v {\rm K})}_{\omega} ({\bf r}) e^{-i\omega t}$ in Eqn.~(\ref{eq-A25}) to obtain the equation,
\begin{equation}
\left( \grad^2 - {\omega^2 \over v^2} \right)  \Phi^{(v {\rm K})}_{\omega} ({\bf r}) = -4 \pi  \rho_{\omega} ({\bf r}).
\label{eq-A28}
\end{equation}
Hence the scalar potential has the solution that vanishes at infinity for both positive and negative $\omega$ and $v$,
\begin{equation}
\Phi^{(v {\rm K})}_{\omega} ({\bf r}, t) = \Phi^{(v {\rm K})}_{\omega} ({\bf r}) e^{-i \omega t}  
= e^{-i \omega t} \int {e^{- |\omega / v|R} \over R} \rho_{\omega} ({\bf r}') d^3 r',
\label{eq-A29}
\end{equation}
where $R = |{\bf r} - {\bf r}'|$.  One recalls that Eqn.~(\ref{eq-A29}) is of the form of an instantaneous Yukawa potential.  If we use the continuity equation, 
$i \omega \rho_\omega ({\bf r}') = \grad' \cdot {\bf J}_\omega ({\bf r}')$, then
\begin{eqnarray}
- c\grad \int^t \Phi^{(v {\rm K})}_{\omega} ({\bf r}, t'') dt'' 
= {c \over (i\omega)^2} e^{-i \omega t} \grad \int {e^{- |\omega / v|R} \over R} [\grad' \cdot {\bf J}_{\omega} ({\bf r}')] d^3 r'
\qquad
\nonumber
\\
= - {1 \over c} {c^2 \over \omega^2} e^{-i \omega t} \int \left[\grad \grad \left( {e^{- |\omega / v|R} \over R} \right)\right]
   \cdot {\bf J}_\omega ({\bf r}') d^3 r'.
\label{eq-A30}
\end{eqnarray}
Thus, the Green's function for the vector potential in the frequency domain is:
\begin{equation}
\bar{{\bf G}}_\omega^{(v {\rm K})}({\bf r} || {\bf r}') 
= {e^{i(\omega / c)R} \over R} \bar{{\bf I}}
+ {c^2 \over \omega^2} \grad \grad \left( {e^{i(\omega / c)R} \over R} -  {e^{-|\omega / v|R} \over R}  \right).
\label{eq-A31}
\end{equation}
In the limit that $v$ goes to infinity, Eqn. (\ref{eq-A31}) reduces to the Coulomb-gauge Green's function in Eqn. (\ref{eq-A18}),
\begin{equation}
\lim_{v \to \infty} \bar{{\bf G}}_\omega^{(v {\rm K})}({\bf r} || {\bf r}') 
= {e^{i(\omega / c)R} \over R} \bar{{\bf I}}
+ {c^2 \over \omega^2} \grad \grad \left( {e^{i(\omega / c)R} \over R} -  {1 \over R}  \right)
= \bar{{\bf G}}_\omega^{(C)}({\bf r} || {\bf r}').
\label{eq-A32}
\end{equation}
In the limit that $v$ goes to zero, Eqn. (\ref{eq-A31}) reduces to the temporal-gauge Green's function in Eqn. (\ref{eq-A23}) for all $\omega \ne 0$,
\begin{equation}
\lim_{v \to 0} \bar{{\bf G}}_\omega^{(v {\rm K})}({\bf r} || {\bf r}') 
= {e^{i(\omega / c)R} \over R} \bar{{\bf I}}
+ {c^2 \over \omega^2} \grad \grad \left( {e^{i(\omega / c)R} \over R} \right) 
= \bar{{\bf G}}_\omega^{(T)}({\bf r} || {\bf r}').
\label{eq-A33}
\end{equation}

We have derived the complete time and frequency domain dyadic Green's functions for the velocity and the temporal gauges and confirmed that the result for the velocity gauge in Ref.~\cite{Yang-gauge-2005} reduces to that of the Lorenz gauge and to the Coulomb gauge result in Ref.~\cite{Nevels-Crowell-1990} as special cases. 
We have also derived the frequency domain Green's function for the generalized Kirchhoff gauge and showed that the Coulomb gauge and temporal gauge results are special cases of the Kirchhoff gauge. 
As is clear from Eqn. (\ref{eq-A33}), the generalized Kirchhoff gauge reduces to the temporal gauge only when there is no zero-frequency component of the charge density.

\section{Conclusions}

In this paper we have derived the dyadic Green's functions for the velocity, temporal and generalized Kirchhoff gauges. For the special cases $v = c$ and $v = \infty$ the velocity gauge Green's function was shown to be in complete agreement with the previously known frequency domain Lorenz and Coulomb gauge dyadic Green's functions.  When $v = \infty$ and $v = 0$, the frequency domain Green's function of the generalized Kirchhoff gauge reduces to the Green's functions for the Coulomb and temporal gauges.  A useful aspect of the generalized velocity and Kirchhoff gauge Green's functions is that by simply adjusting the parameter $v$ one can obtain Green's function expressions for an infinite continuous set of gauges.

\appendix
\section{Appendix: Scalar potential of the generalized Kirchhoff gauge}
\setcounter{equation}{0}
\renewcommand{\theequation}{A.\arabic{equation}}

Heras~\cite{Heras-2006} has shown that the scalar potential with an imaginary propagating speed $\pm iv$,
\begin{equation}
\Phi^{(v {\rm K})} ({\bf r}, t) = \int {1 \over R} \rho({\bf r}', t - R/(\pm iv)) d^3r',
\qquad \qquad 
R = |{\bf r} - {\bf r}'|,
\label{eq-ap1}
\end{equation}
is a formal solution of Eqn.~(\ref{eq-A25}).  He only considered the original
Kirchhoff gauge with $v = c$. But his results can be generalized to an arbitrary real $v \ne 0$. 
Unfortunately, the boundary condition that the scalar potential vanishes at infinity makes it difficult to apply Eqn.~(\ref{eq-ap1}) to all forms of time variation in a straightforward manner.   Consider a charge density of the form,
\begin{equation}
\rho({\bf r}, t) = 2 \rho_\omega ({\bf r}) cos (\omega t),
\label{eq-ap2}
\end{equation}
with positive $\omega$ and real $\rho_\omega({\bf r})$.  Let us assume that $v$ is positive.  If we choouse $+iv$, we get
\begin{eqnarray}
\Phi^{(v {\rm K})}_{+iv}({\bf r}, t) = \int {1 \over R} \rho({\bf r}', t - R/(+ iv)) d^3r'
\qquad \qquad \qquad \qquad \qquad \qquad 
\nonumber
\\
= e^{i \omega t} \int {1 \over R} \rho_\omega ({\bf r}') e^{-\omega R/v} d^3r'
  + e^{- i \omega t} \int {1 \over R} \rho_\omega ({\bf r}') e^{\omega R/v} d^3r',
\label{eq-ap3}
\end{eqnarray}
which does not vanish at infinity in space.  Similarly, if we choose $-iv$ in eqn.~(\ref{eq-ap1}), we get
\begin{eqnarray}
\Phi^{(v {\rm {\rm K}})}_{-iv}({\bf r}, t) = \int {1 \over R} \rho({\bf r}', t - R/(- iv)) d^3r'
\qquad \qquad \qquad \qquad \qquad \qquad 
\nonumber
\\
= e^{i \omega t} \int {1 \over R} \rho_\omega ({\bf r}') e^{\omega R/v} d^3r'
  + e^{- i \omega t} \int {1 \over R} \rho_\omega ({\bf r}') e^{-\omega R/v} d^3r',
\label{eq-ap4}
\end{eqnarray}
which has he same problem of not vanishing at spatial infinity.  But if we choose $+iv$ for $e^{i \omega t}$ and $-iv$ for $e^{-i \omega t}$, the desired solution is obtained:
\begin{eqnarray}
\Phi^{(\nu {\rm K})}({\bf r}, t) 
= e^{i \omega t} \int {1 \over R} \rho_\omega ({\bf r}') e^{-\omega R/\nu} d^3r'
  + e^{- i \omega t} \int {1 \over R} \rho_\omega ({\bf r}') e^{-\omega R/\nu} d^3r'
\nonumber
\\
= \int {e^{-\omega R/\nu} \over R} \delta(t - t') \cdot 2 \rho_{\omega}({\bf r}') cos(\omega t') d^3r' dt'.
\qquad \quad \quad \quad \ \ 
\label{eq-ap5}
\end{eqnarray}
This scalar potential has the form of an instantaneous Yukawa potential.

For a general charge density we write its complete Fourier integral~\cite{note-Fourier}
\begin{equation}
\rho({\bf r}, t) = \int_{-\infty}^{+\infty} \rho_\omega({\bf r}) e^{-i \omega t} d\omega,
\label{eq-ap6}
\end{equation}
with $\rho_{-\omega} ({\bf r}) = [\rho_\omega ({\bf r})]^*$.  Then the scalar potential for both positive and negative $\omega$ and 
$v$ is:
\begin{equation}
\Phi^{(v {\rm K})}({\bf r}, t) = \int_{- \infty}^{+\infty} d\omega \, e^{-i \omega t} \int {e^{-|\omega / v|R} \over R} 
    \rho_\omega ({\bf r}') 
d^3 r'.
\label{eq-ap7}
\end{equation}
We now prove that the $\Phi^{(v {\rm K})}$ in eq.~(\ref{eq-ap7}) is a solution to eq.~(\ref{eq-A25}), the equation for the scalar potential in the generalized Kirchhoff gauge:
\begin{eqnarray}
\left( \grad^2 + {1 \over v^2} {\partial^2 \over \partial t^2} \right) \Phi^{(v {\rm K})}({\bf r}, t)
= \int_{- \infty}^{+\infty} d\omega \, e^{-i \omega t}  \int \left[ \left( \grad^2 - {\omega^2 \over v^2} \right)
   \left( {e^{-|\omega / v|R} \over R} \right) \right] \rho_\omega ({\bf r}') d^3r'
\nonumber
\\
= \int_{- \infty}^{+\infty} d\omega \, e^{-i \omega t}  \int [ - 4\pi \delta({\bf r} - {\bf r}')] \rho_\omega ({\bf r}') d^3r'
= - 4\pi \int_{- \infty}^{+\infty}  \rho_\omega ({\bf r})e^{-i \omega t} d\omega 
= - 4 \pi \rho ({\bf r}, t).
\ 
\end{eqnarray}
This completes our discussion of the scalar potential of the generalized Kirchhoff gauge.


\begin{thebibliography}{99}
\bibitem{note-Fourier}
We use the following convention for the Fourier integrals for the charge density $\rho({\bf r}, t)$,
the current density ${\bf J}({\bf r}, t)$, the scalar potentials $\Phi({\bf r}, t)$ and the vector potentials ${\bf A}({\bf r}, t)$:
\begin{equation}
\rho_{\omega}({\bf r}) = {1 \over 2\pi} \int_{- \infty}^{+ \infty} \rho ({\bf r}, t) e^{i \omega t}dt, 
\qquad \qquad
\rho ({\bf r}, t) = \int_{- \infty}^{+ \infty} \rho_{\omega}({\bf r}) e^{- i \omega t}d \omega.
\nonumber
\end{equation}
\end{thebibliography}
\end{document}